\documentstyle[preprint,aps,epsfig]{revtex}
\tightenlines
\begin{document}
\draft 
\preprint{\vbox{\hbox{hep-ph/0001030}
\hbox{FTUV/00-01}
\hbox{IFIC/00-01}
\hbox{IFT-P.004/2000}
\hbox{LBNL-44767}}}

\title{Probing Intermediate Mass Higgs Interactions \\
       at the CERN Large Hadron Collider}
\author{O.\ J.\ P.\ \'Eboli  $^{1}$,
        M.\ C.\ Gonzalez--Garcia $^{2}$,
        S.\ M.\ Lietti $^3$, and
        S.\ F.\ Novaes $^1$}
\address{
$^1$ Instituto de F\'{\i}sica Te\'orica,
     Universidade  Estadual Paulista, \\
     Rua Pamplona 145, 01405--900, S\~ao Paulo, Brazil\\
$^2$ Instituto de F\'{\i}sica Corpuscular IFIC CSIC--Universidad de Valencia,\\
     Edificio Institutos de Paterna, Apartado 2085, 46071 Valencia\\    
$^3$ Lawrence Berkeley National Laboratory \\
     Berkeley, CA 94720, USA }
\date{December 23, 1999}
\maketitle
\begin{abstract}
\vskip -36pt
We analyze the potentiality of the CERN Large Hadron Collider  to
probe the Higgs boson couplings to the electroweak gauge bosons.  We
parametrize the possible deviations of these couplings due to new
physics in a model independent way, using the most general
dimension--six effective lagrangian where the $SU(2)_L \otimes
U(1)_Y$ is realized linearly. For intermediate Higgs masses, the
decay channel into two photons is the most important one for Higgs
searches at the LHC. We study the effects of these new interactions
on the Higgs production mechanism and its subsequent decay into two
photons. We show that the LHC will be sensitive to new physics scales
beyond the present limits extracted from the LEP and Tevatron
physics.
\end{abstract}
\newpage
\section{Introduction}
\label{int}

The search for the Higgs boson and the study of the $SU(2)_L \otimes
U(1)_Y$ symmetry breaking mechanism are the main goals of the present
and future high energy experiments \cite{review}.  At present, the
best available limit on the Higgs mass arises from searches at the
CERN LEP collider. The ALEPH Collaboration analysis of the 1999 data
with integrated luminosities of 29 pb$^{-1}$ at $\sqrt{s}=191.6$ GeV
and 69.5 pb$^{-1}$ at $\sqrt{s}=195.6$ GeV \cite{aleph} yields
$M_H>98.8$ GeV at 95\% CL. A next step will be given by the CERN
Large Hadron Collider (LHC) that will be able to detect the standard
Higgs boson in the decay channel $H \rightarrow \gamma\gamma$ for
masses in the range of 100--150 GeV \cite{atlas,cms}.

If a Higgs boson is observed then it is imperative to verify whether
its couplings are in agreement with the ones predicted by the
Standard Model (SM) with a single scalar doublet. In the SM, the
precise form of the Higgs couplings to the gauge bosons and its
self--couplings are completely determined in terms of one free
parameter which can be chosen to be the Higgs mass $M_{H}$.  However,
in this simple realization, the theory presents  the problem that
the quantum corrections to the Higgs mass are quadratically divergent
with the cut--off.  This implies the necessity of a large
fine--tuning in order to keep the theory perturbative up to very high
energies, or, conversely, the existence of new physics which manifest
itself above a certain scale $\Lambda$.  If the energy scale of new
physics is large compared to the electroweak scale and there is no
new light resonances, we can represent its impact on Higgs boson
properties via the introduction of effective operators \cite{concha}.
In this approach, we parametrize the Higgs anomalous interactions by
the most general dimension--six effective lagrangian with the linear
realization of the $SU(2)_L \otimes U(1)_Y$ symmetry.

In the linear realization of the SM there are eleven $C$ and $P$
conserving dimension--six operators with some of them contributing at
tree level to well measured observables, and consequently being
severely constrained \cite{rujula,dim6:zep}. In Ref.\ \cite{rujula}
it was argued that it is unnatural to expect a large hierarchy
between the coefficients of the various dimension--six operators
independently on whether they do or do not contribute at tree level
to the low energy observables.  As a consequence the existing limits
coming from the LEP I physics would preclude the direct observation
of new effects in the processes accessible at LEP II and Tevatron.

In this work, we study the potentiality of the CERN LHC to probe the
interactions of an intermediate mass Higgs boson and consequently its
sensitivity to new physics scales beyond the constraints stemming
from the tree level contribution of the effective operators to low
energy observables. For intermediate Higgs masses, the decay channel
$H \rightarrow \gamma\gamma$ is the most important one for the Higgs
search at the LHC. Moreover the effect of the additional Higgs
interactions on the properties of an intermediate mass Higgs can be
more easily seen in processes that are suppressed in the SM, such as 
the Higgs decay into two photons. In the SM, this decay occurs only
at one--loop level, and it can be enhanced (or suppressed) by the
anomalous interactions.

We analyze the Higgs production and subsequent decay into two photons
through gluon--gluon fusion
\begin{equation}
pp ~\to ~g g ~\to~ H ~(\to~ \gamma \gamma) \;\; ,
\label{fus:gg}
\end{equation}
as well as through vector--boson fusion mechanism
\begin{equation}
pp ~\to~ q q^\prime  V V ~\to~ j + j + H (\to \gamma \gamma) \;\; .
\label{fus:ww}
\end{equation}
with $V= W^\pm$ or $Z^0$. We show that the LHC will be able to expand
considerably the present sensitivity on the dimension--six Higgs
couplings, being able to probe new physics scales as large as 2.2
TeV, provided that the Higgs is observed.  In this case, our results
show that the LHC will be able to improve the limits arising from the
tree level contribution to the LEP I observables, which are presently
the most severe constraint.  Furthermore, there is also a distinct
possibility, {\it i.e.} the existence of anomalous Higgs interactions
reduces its decay into two photons. In this case, no signal will be
observed in the above reactions, and, consequently, if the Higgs is
observed in other decay channel, the existence of non--vanishing
anomalous couplings could also be established.

\section{Effective Higgs Interactions and Present Bounds}

In the linear representation of the $SU(2)_L \otimes U(1)_Y$ symmetry
breaking mechanism, the SM model is the lowest order approximation
while the first corrections, which are of dimension six, can be
written as
\begin{equation}
{\cal L}_{\text{eff}} = \sum_n \frac{f_n}{\Lambda^2} {\cal O}_n \;\; ,
\label{l:eff}
\end{equation}
where the operators ${\cal O}_n$ involve vector--boson and/or
Higgs--boson fields with couplings $f_n$ \cite{linear}. This
effective Lagrangian describes the phenomenology of models that
are somehow close to the SM since a light Higgs scalar doublet is
still present at low energies. Of the eleven possible operators
${\cal O}_{n}$ that are $P$ and $C$ even, only seven of them modify
the Higgs--boson couplings to vector bosons
\cite{dim6:zep,Hagiwara2},
\begin{eqnarray}
&&{\cal O}_{BW} =  \Phi^{\dagger} \hat{B}_{\mu \nu} 
\hat{W}^{\mu \nu} \Phi \;\; , \nonumber \\ 
&&{\cal O}_{WW} = \Phi^{\dagger} \hat{W}_{\mu \nu} 
\hat{W}^{\mu \nu} \Phi  \;\; , \nonumber \\
&&{\cal O}_{BB} = \Phi^{\dagger} \hat{B}_{\mu \nu} 
\hat{B}^{\mu \nu} \Phi \;\; ,  
\nonumber 
\\
&&{\cal O}_W  = (D_{\mu} \Phi)^{\dagger} 
\hat{W}^{\mu \nu}  (D_{\nu} \Phi) \;\; , 
\label{eff}  \\
&&{\cal O}_B  =  (D_{\mu} \Phi)^{\dagger} 
\hat{B}^{\mu \nu}  (D_{\nu} \Phi)  \;\; , \nonumber \\
&&{\cal O}_{\Phi,1} = \left ( D_\mu \Phi \right)^\dagger \Phi^\dagger \Phi
\left ( D^\mu \Phi \right ) \;\; , \nonumber \\
&&{\cal O}_{\Phi,2} = \frac{1}{2} 
\partial^\mu\left ( \Phi^\dagger \Phi \right)
\partial_\mu\left ( \Phi^\dagger \Phi \right)
 \;\; , \nonumber
\end{eqnarray}
where $\Phi$ is the Higgs doublet, $D_\mu$ the covariant derivative,
$\hat{B}_{\mu \nu} = i (g'/2) B_{\mu \nu}$, and $\hat{W}_{\mu \nu} =
i (g/2) \sigma^a W^a_{\mu \nu}$, with $B_{\mu \nu}$ and $ W^a_{\mu
\nu}$ being respectively the $U(1)_Y$ and $SU(2)_L$ field strength
tensors. It is interesting to notice that the operators ${\cal
O}_{\Phi,1}$ and ${\cal O}_{\Phi,2}$ contribute to the weak boson
mass and Higgs wave function, which in turn leads to new Higgs
couplings to the gauge bosons.

The effective operators in Eq.\ (\ref{eff}) give rise to anomalous
$H\gamma\gamma$, $HZ\gamma$, $HZZ$, and $HW^+W^-$ couplings, which,
in the unitary gauge, are given by
\begin{eqnarray}
{\cal L}_{\text{eff}}^{\text{HVV}} &=& 
g_{H \gamma \gamma} \; H A_{\mu \nu} A^{\mu \nu} + 
g^{(1)}_{H Z \gamma} \; A_{\mu \nu} Z^{\mu} \partial^{\nu} H + 
g^{(2)}_{H Z \gamma} \; H A_{\mu \nu} Z^{\mu \nu}
\nonumber \\
&+& g^{(1)}_{H Z Z}  \; Z_{\mu \nu} Z^{\mu} \partial^{\nu} H + 
g^{(2)}_{H Z Z}  \; H Z_{\mu \nu} Z^{\mu \nu} +
h^{(3)}_{H Z Z}  \; H Z_\mu Z^\mu  \\
\label{eff:nn}
&+& g^{(1)}_{H W W}  \; \left (W^+_{\mu \nu} W^{- \, \mu} \partial^{\nu} H 
+\text{h.c.} \right) +
g^{(2)}_{H W W}  \; H W^+_{\mu \nu} W^{- \, \mu \nu}
+g^{(3)}_{H W W}  \; H W^+_{\mu} W^{- \, \mu}
 \;\; ,\nonumber
\end{eqnarray}
where $A(Z)_{\mu \nu} = \partial_\mu A(Z)_\nu - \partial_\nu
A(Z)_\mu$. The effective couplings $g_{H \gamma \gamma}$,
$g^{(1,2)}_{H Z \gamma}$, and $g^{(1,2,3)}_{H Z Z}$ are related
to the coefficients of the operators appearing in (\ref{l:eff})
through,
\begin{eqnarray}
g_{H \gamma \gamma} &=& - \left( \frac{g M_W}{\Lambda^2} \right)
                       \frac{s^2 (f_{BB} + f_{WW} - f_{BW})}{2} \;\; , 
\nonumber \\
g^{(1)}_{H Z \gamma} &=& \left( \frac{g M_W}{\Lambda^2} \right) 
                     \frac{s (f_W - f_B) }{2 c} \;\; ,  
\nonumber \\
g^{(2)}_{H Z \gamma} &=& \left( \frac{g M_W}{\Lambda^2} \right) 
                      \frac{s [2 s^2 f_{BB} - 2 c^2 f_{WW} + 
                     (c^2-s^2)f_{BW} ]}{2 c}  \;\; , 
\nonumber \\ 
g^{(1)}_{H Z Z} &=& \left( \frac{g M_W}{\Lambda^2} \right) 
	              \frac{c^2 f_W + s^2 f_B}{2 c^2} \nonumber \;\; , \\
g^{(2)}_{H Z Z} &=& - \left( \frac{g M_W}{\Lambda^2} \right) 
  \frac{s^4 f_{BB} +c^4 f_{WW} + c^2 s^2 f_{BW}}{2 c^2} \label{g} \;\; , \\
g^{(3)}_{H Z Z} &=& \left( \frac{ g M_W v^2}{\Lambda^2} \right) 
	              \frac{f_{\Phi,1}-f_{\Phi,2}}{4 c^2} \;\; , \nonumber \\
g^{(1)}_{H W W} &=& \left( \frac{g M_W}{\Lambda^2} \right) 
                      \frac{f_{W}}{2}  \;\; , \nonumber \\
g^{(2)}_{H W W} &=& - \left( \frac{g M_W }{\Lambda^2} \right) 
  f_{WW}  \;\; , \nonumber \\
g^{(3)}_{H W W} 
&=& -  \left( \frac{ g M_W v^2}{\Lambda^2} \right) 
	              \frac{f_{\Phi,1}+2 f_{\Phi,2}}{4}  \;\; ,  
\nonumber
\end{eqnarray}
with $g$ being the electroweak coupling constant and $s(c) \equiv
\sin(\cos)\theta_W$. In the couplings $g^{(3)}_{H Z Z}$ and
$g^{(3)}_{H W W}$ we have also included the effects arising from the
contribution of the operators ${\cal O}_{\Phi,1}$ and ${\cal
O}_{\Phi,2}$ to the renormalization of the weak boson masses and the
Higgs field wave function.

The operators ${\cal O}_{\Phi,1}$ and ${\cal O}_{BW}$ contribute at
tree level to the vector--boson two--point functions, and
consequently are severely constrained by low--energy data
\cite{rujula,dim6:zep}. The present 95\% CL limits on these operators
for 90 GeV $ \le M_H \le$ 800 GeV and $m_{\text{top}}=175$ GeV read
\cite{hms},
\begin{eqnarray}
-1.2  \le &&\frac{f_{\Phi,1}}{\Lambda^2} \le 0.56 \hbox{ TeV}^{-2} 
\;\; , 
\label{b:phi1} \\
-1.0 \le &&\frac{f_{BW}}{\Lambda^2} \le 8.6 \hbox{ TeV}^{-2} \;\; .
\label{b:bw}
\end{eqnarray}

On the order hand, the remaining operators can be indirectly
constrained via their one--loop contributions to low--energy
observables, which are suppressed by factors $1/(16 \pi^2)$.
Using the ``naturalness'' assumption that large cancellations do not
occur among their contributions, we can consider only the effect of
one operator at a time. In this case, the following
constraints at 95\% CL (in units of TeV$^{-2}$) arise \cite{hms}
\begin{eqnarray}
&& -12.\leq \frac{f_{W}}{\Lambda^2}\leq 2.5\; ,  \nonumber\\
&& -7.6\leq \frac{f_{B}}{\Lambda^2}\leq 22\; ,  \label{blindlim}\\
&& -24\leq \frac{f_{WW}}{\Lambda^2}\leq 14\; ,  \nonumber\\
&& -79\leq \frac{f_{BB}}{\Lambda^2}\leq 47\; .  \nonumber
\end{eqnarray}
These limits depend in a complex way on the Higgs mass. The values
quoted above for the sake of illustration were obtained for 
$M_{H}=200$ GeV.

Some of the anomalous Higgs interaction can also be constrained by
their effect on the triple gauge--boson vertices. Recently, the LEP
and Tevatron Collaborations have studied the production of pairs of
gauge bosons and derived bounds on the anomalous interactions that
modify the $WW\gamma$ and $WWZ$ vertices. Combining the published
results from D\O\ and the four LEP experiments, the 95\% CL bounds on
the anomalous Higgs interactions are (in TeV$^{-2}$) \cite{lepd0}:
\begin{eqnarray}
&&-31\leq \frac{(f_W+f_B)}{\Lambda^2}\leq 68 \;, \;\;\; 
\mbox{ for } \;\;\;  f_{WWW} = 0 \;\; .
\label{b:wwv}
\end{eqnarray}
Notice that, since neither $f_{WW}$ nor $f_{BB}$ contribute to the
triple gauge--boson vertices, no direct constraint on these couplings
can be derived from this analysis. Notwithstanding, these couplings
can be constrained by data on Higgs searches at LEP II
\cite{ours:lep2aaa} and Tevatron \cite{ours:tevatron} colliders.  The
combined analysis \cite{ours:comb} of these signatures yields the
following 95\% CL bounds on the anomalous Higgs interactions (in
TeV$^{-2}$):
\[
-7.5 \leq  \frac{f_{WW(BB)}}{\Lambda^2} \leq 18
\]
for $M_H\leq 150$ GeV. These limits can be improved by a factor 2--3
in the upgraded Tevatron runs.


\section{Intermediate Mass Higgs Production and New Interactions}

\subsection{Gluon--gluon fusion}

At the LHC, light Higgs bosons are copiously produced through
gluon-gluon fusion and it was established that the decay mode $H \to
\gamma\gamma$ provides the best signature for Higgs masses in the
range $90 < M_H < 150$ GeV \cite{h:rev}.  However, this channel also 
possesses a very large background, requiring a good energy resolution
from the detectors in order to be observed. In our analyses we
computed the SM and anomalous Higgs production cross sections using
the cuts and efficiencies that the ATLAS Collaboration applied in
their studies. We imposed the acceptance cuts
\[
p_{T}^{\gamma_{1(2)}}  >  25~ (40) \; \text{GeV} \;\;\; , 
\;\;\;\;\;\;	
|\eta_{\gamma_{(1,2)}}| <   2.5 	\;\; , 	
\]
where $p_{T}^{\gamma_{1(2)}}$ stands for the largest (second largest)
transverse momentum of the photons in the event and
$\eta_{\gamma_{(1,2)}}$ are the rapidities of these photons. We also
required that
\[
\frac{p_{T}^{\gamma_1}}{(p_{T}^{\gamma_1} + p_{T}^{\gamma_2})}  <   0.7
\]
to reduce the quark bremsstrahlung background. The efficiency for
reconstruction and identification of one photon was taken to be
80\%.  We collected the photon pair events whose invariant masses
fall in bins of  size $2\Delta M_H$ around the Higgs mass with
\[
\left( \frac{\Delta M_H}{M_H} \right) =  \frac{5\%}{\sqrt{M_H}} 
\oplus 0.5\% 
\;\; .
\]

In order to take advantage of the careful estimates of backgrounds
and detector efficiencies made by the ATLAS collaboration, we
normalized our predictions for the SM Higgs cross section to their
value for each value of the Higgs mass and then we rescaled the
anomalous production cross section by the same factor. This factor
varies from 0.65 to 0.80, depending on the Higgs mass. We present in
Table \ref{gg:evt} the expected number of Higgs and background events
in the SM for a center of mass energy $\sqrt{s}=14$ TeV and 
an integrated luminosity of 100 fb$^{-1}$ after cuts and
efficiencies. The SM irreducible background is by far the most
important one, however, there is still rather large jet--jet and
$\gamma$--jet reducible backgrounds.  It is worth mentioning that the
SM Higgs can be observed in this channel with an statistical
significance of more than 3.9 standard deviations.

\begin{figure}
\begin{center}
\mbox{\epsfig{file=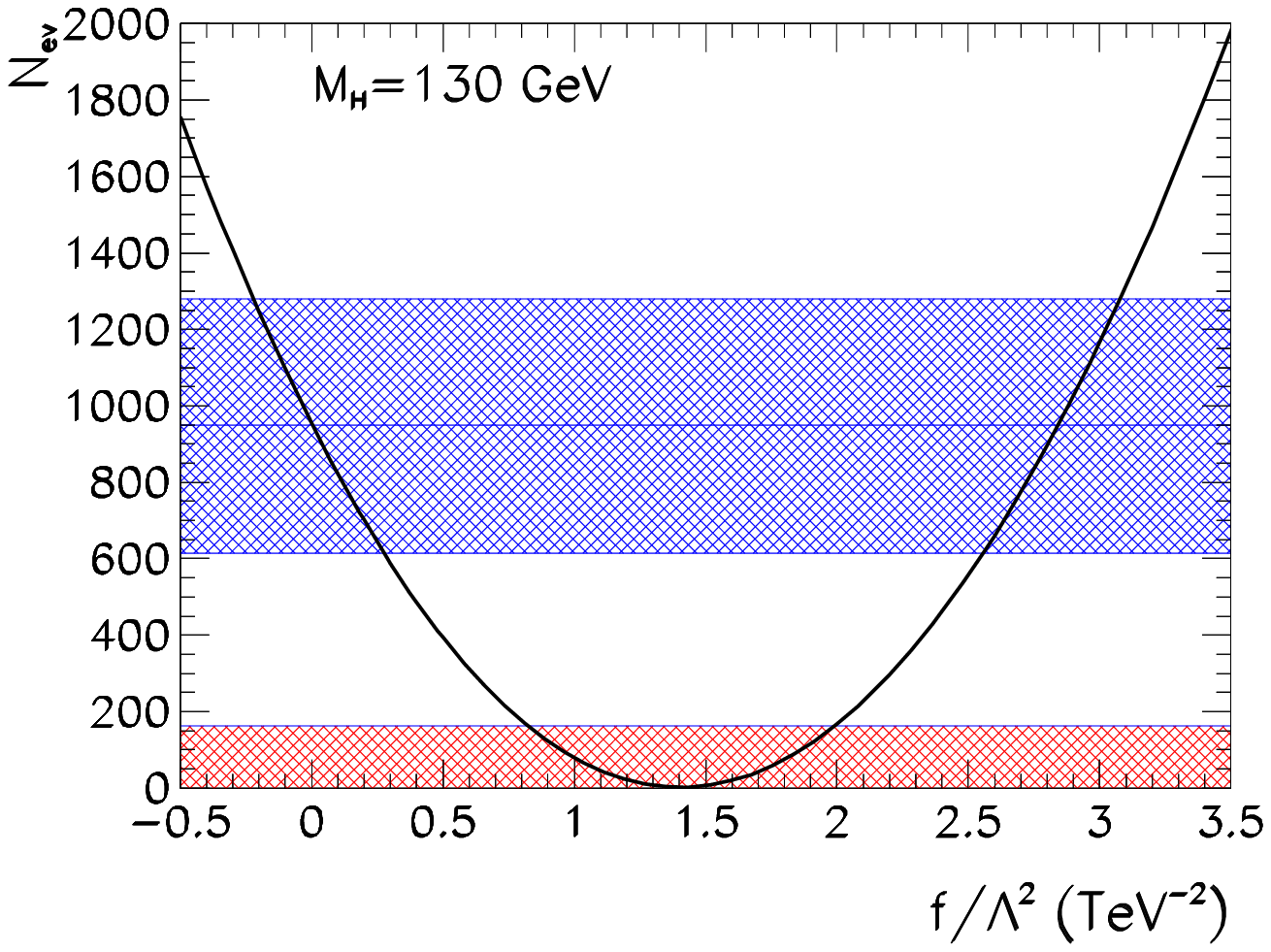,width=0.8\textwidth}}
\end{center} 
\caption{Number of events $pp ~\to ~g g ~\to~ H ~\to~ \gamma \gamma$
as a function of the anomalous coupling $f$ [Eq.({\protect\ref{f}})] for
$M_H$=130 GeV and an integrated luminosity of 100 fb$^{-1}$. We  show
the 95\% CL allowed region assuming that the SM Higgs is observed
(upper shadowed area). The lower shadowed area represents the 95\% CL
allowed region for the scenario where only the SM background is
observed, {\em i.e.} with no SM Higgs signal is seen.}
\label{fig:ggh} 
\end{figure}

For an intermediate mass Higgs, ${\cal O}_{BB}$, ${\cal O}_{WW}$, and
${\cal O}_{BW}$ are the only effective operators (\ref{eff})
that contribute significantly at tree level to the process $gg \to H
\to \gamma \gamma $. For the sake of illustration, we exhibit in
Fig.\ \ref{fig:ggh} the expected number of events after cuts for a
Higgs boson with $M_H=130$ GeV as a function of the anomalous
coupling 
\begin{equation}
f \equiv \frac{(f_{BB} + f_{WW} - f_{BW})}{\Lambda^2} \; . 
\label{f}
\end{equation}
Notice that the anomalous contribution to $H \to \gamma \gamma$
depends only on this combination of the effective interactions [see
Eq.\ (\ref{g})]. 

Let us initially assume that the Higgs boson has indeed been observed
in the $\gamma\gamma$ channel and that its production cross section
is in agreement with the SM prediction. In this case we can extract
the sensitivity of this process to the anomalous interactions using
as background the processes listed in Table \ref{gg:evt} together
with the SM Higgs signal. We present in Table \ref{gg:h1} the 95\% CL
allowed range of $f$ for an integrated luminosity of 100 fb$^{-1}$.
In Fig.\ \ref{fig:ggh}, we also exhibit the 95\% CL interval around
the SM Higgs signal (upper shadowed region) for $M_H=130$ GeV.  The
two allowed parameter ranges given in Table \ref{gg:h1} correspond to
the two intersections of the signal curve with the 95\% CL region. 
One should notice that the existence of a sizable interference
between the SM and anomalous contributions to the Higgs into two
photons width allows the existence of a range of anomalous couplings
not compatible with zero even if its production rate is in accordance
with the SM predictions. This range would correspond to $2.5 <
f/\Lambda^2 < 3.0$ TeV$^{-2}$, for $M_H = 130$ GeV (see Fig.\
\ref{fig:ggh}). 

The bounds in Table \ref{gg:h1}, derived from the SM Higgs
observation, are more restrictive than the ones emanating from the
contribution of these operators to the low--energy and LEP physics
either at tree level (\ref{b:bw}) or at the one--loop level 
(\ref{blindlim}).  As seen in this table for any Higgs mass, this
process is sensitive to new physics which characteristic scale up to
$ \Lambda\simeq 2.2$ TeV for $f_{BB} = f_{WW} = f_{BW}=1$.

Another possible scenario is the one where no SM Higgs signal is
observed in the $\gamma \gamma$ channel. In this case, we obtain
different constraints on the anomalous couplings for a given value of
$M_H$. In order to extract the sensitivity region for the anomalous
couplings we have to consider only the SM Higgs backgrounds listed in
Table \ref{gg:evt} without including in that the SM Higgs signal.  In
Table \ref{gg:h2} we present sensitivity range for the anomalous
coupling $f$ at 95\% CL, assuming that only the SM Higgs background
is observed for an integrated luminosity of 100 fb$^{-1}$.  The
interpretation of these results depend on the observation of the
Higgs in other channels. Should the Higgs be observed in another
decay channel insensitive to these anomalous couplings, such as in
$H\rightarrow\tau^+\tau^-$ \cite{tautau}, the non observation of the
corresponding signal in $\gamma\gamma$ should be interpreted as the
inevitable existence of new physics in the Higgs couplings to photons
with characteristic strength as given in Table \ref{gg:h2}.  If, on
the other hand, no signal is observed in any other decay channel, the
existence of the Higgs of a given mass can be ruled out and no limit
can be extracted from the non observation of the $\gamma\gamma$
signal, becoming meaningless the constraints in Table \ref{gg:h2}.


\subsection{Vector boson fusion}

Higgs bosons can also be produced in $pp$ collisions via the vector
boson fusion (VBF) process (\ref{fus:ww}).  A nice feature of this
production mechanism is that the jets in the VBF signal events tend
to populate the forward direction and can be used to tag these events
and reduce the background. Moreover, the background can be further
suppressed by vetoing additional jet activity in the central region
\cite{bpz}.

In our analysis of this process we followed closely the study of
Ref.\ \cite{die}.  We considered all backgrounds from QCD and
electroweak processes, as well as the corresponding interferences,
which can lead to events with two photons and two jets.  We generated
the scattering amplitudes for both the SM Higgs signal and the 
background with the package Madgraph~\cite{madgraph} which makes use
of the helicity amplitudes contained in the package Helas
\cite{helas}. In our calculations we have used the MRS(G) \cite{mrs}
proton structure function.

In order to reduce the background, we required the photons to satisfy 
the acceptance cuts
\[
 p_{T}^{\gamma_{1(2)}} >  25~ (50) \; \text{GeV} \;\;\;  ,\;\;\;\;\;\;
|\eta_{\gamma_{(1,2)}}|  <   2.5 \;\; ,
\]
and we tagged the jets in the forward region and vetoed their
presence in the central detector
\begin{eqnarray*}
&&p_{T}^{j_{(1,2)}} >  20~ (40)  \; \text{GeV} \;\; , \\
|\eta_{j_{(1,2)}}| 	<  5.0  \;\; , \;\;\;\;
&&|\eta_{j_{1}} - \eta_{j_{2}}|  >  4.4 \;\; ,\;\;\;\;
\eta_{j_{1}} . \eta_{j_{2}}  < 0  \;\; .
\end{eqnarray*}
We also imposed that the photons are isolated from the jets ($\Delta
R_{\gamma j}> 0.7 $) and, as before, used the photon detection
efficiency of 80\%.

We display in Table \ref{vbf:evt} the expected number of SM Higgs
signal and background events using the above cuts and efficiencies
for an integrated luminosity of 100 fb$^{-1}$. Our results show good
quantitative agreement with the analysis in Ref.\ \cite{die}, taking
into account the different choice of structure functions and the
inclusion of the photon detection efficiency. Notice that the SM
irreducible background and the SM Higgs signal are of the same order
of magnitude.

The dimension--six operators in Eq.\ (\ref{eff}) contribute to the
$\gamma\gamma$ signal of Higgs production via vector boson fusion by
modifying both the production cross section as well as the Higgs
decay width.  On one hand, the operators ${\cal O}_{BB}$, ${\cal
O}_{WW}$, and ${\cal O}_{BW}$ contribute to the Higgs production and
decay while ${\cal O}_{W}$, ${\cal O}_{B}$, ${\cal O}_{\Phi,1}$, and
${\cal O}_{\phi,2}$ change only the production vertices. In our
calculation of the anomalous contribution we have included all
amplitudes generated by these operators in the spirit of Refs.\
\cite{ours:lep2aaa,ours:tevatron}.

As in the gluon--gluon fusion analysis, we first assume that the
Higgs boson has indeed been observed in the $\gamma\gamma j j$
channel and that its production cross section is in agreement with
the SM prediction.  In this case, the SM backgrounds for the
anomalous interactions are the processes listed in Table
\ref{vbf:evt} together with the SM Higgs signal.  We present in Table
\ref{vbf:h1} the 95\% CL sensitivity to anomalous coupling
combination (\ref{f}) assuming that the other anomalous couplings
vanish, for an integrated luminosity of 100 fb$^{-1}$. This Table also
contains the 95\% CL allowed range for the ``super--blind'' operator
${\cal O}_{\Phi,2}$ alone as well as the limits attainable assuming
all couplings to be equal $(f_{all} = f_{BB} =f_{WW} =f_{BW} =f_{W}
=f_{B} = f_{\Phi_1} =f_{\Phi_2})$.

Here we also find two allowed ranges of anomalous parameters due to
the sizeable interference between the SM and the anomalous
contributions to the Higgs production and decay.  Comparing the
results in Tables \ref{gg:h1} and \ref{vbf:h1} we see that for those
operators contributing to the Higgs decay into two photons, the VBF
process leads to slightly better sensitivity to new interactions.  On
the other hand, the LHC capability to probe the operators
contributing only to the Higgs production in VBF is much smaller,
being $f_{\Phi,2}$ the only coupling that can be meaningfully
constrained.  For $f_B$, $f_W$, and $f_{\Phi,1}$ we verified that the
expected sensitivity from this analysis is much worse than present
limits in Eqs.\ (\ref{b:phi1}), (\ref{b:bw}), and (\ref{b:wwv}).

Different sensitivity bounds are obtained if no Higgs boson signal is
observed in the VBF channel for a given value of $M_H$.  In this
scenario, we assumed that only the SM background for the Higgs
search  were observed (see Table \ref{vbf:evt}).  In Table
\ref{vbf:h2} we show sensitivity range of anomalous couplings at the
95\% CL for an integrated luminosity of 100 fb$^{-1}$ assuming that
no Higgs signal is observed in the corresponding mass bin.

As discussed above the interpretation of these results depend whether
the Higgs has been observed in other channels. The effective
interactions ${\cal O}_{BB}$, ${\cal O}_{WW}$, and ${\cal O}_{BW}$
can diminish the Higgs decay width into $\gamma\gamma$ and
consequently this Higgs signature will not be observed neither in
gluon--gluon fusion nor in VBF. In this case, the VBF Higgs
production can be seen in the $\tau^+ \tau^-$ channel. If the Higgs
is really discovered in this channel, the analysis of the
$\gamma\gamma$ channel is a strong sign of the existence of new Higgs
interactions. Of course, the constraints on the anomalous couplings
are meaningless if the Higgs is not seen in any channel. For the
operators ${\cal O}_{W}$, ${\cal O}_{B}$, ${\cal O}_{\Phi,1}$, and
${\cal O}_{\phi,2}$ the main effect is to modify the production cross
section, affecting equally the Higgs signal in all decay modes.
Therefore, the bounds on these operators make sense only if the Higgs
production via gluon--gluon fusion is observed.


\section{Conclusions}

In this work we have studied the sensitivity of the LHC collider to
new physics in the Higgs sector of the SM. In particular we have
concentrated on new signals associated to the decay of an
intermediate mass Higgs boson in two photons both in gluon--gluon
fusion (\ref{fus:gg}) as well as in gauge--boson fusion
(\ref{fus:ww}). 

We have shown that the LHC will be able to expand considerably the
present sensitivity on the dimension--six Higgs couplings that modify
the $H\gamma\gamma$ vertex, being able to probe new physics scales as
large as 2.2 TeV, provided that the Higgs is observed.  In both
channels our result show that of the LHC is sensitive to new physics
scales beyond the present constraints originating from their
contribution at tree level to the LEP I and low energy observables.
For the effective operators that do not change the Higgs coupling to
photons, the LHC possible bounds are weaker than the ones presently
available, with the exception of the ``super--blind'' operator ${\cal
O}_{\Phi,2}$.

We have also found that due to the presence of a sizeable
interference between the anomalous and the SM contributions there is
the distinct possibility that the anomalous Higgs interactions dilute
its decay into two photons and the Higgs may be not observable in the
above reactions. Thus the observation of the Higgs in other decay
channel, namely $\tau^+\tau^-$, would imply the existence of new
physics in the Higgs couplings to gauge bosons at a characteristic
scale of ~ 0.7--1.4 TeV.

\section*{Acknowledgments}
M.~C. G.-G. is thankful to the IFT for their kind 
hospitality during her visit.  
This work was supported in part by the Director, Office of Science,
Office of High Energy and Nuclear Physics, Division of High Energy
Physics of the U.S. Department of Energy under Contract 
DE-AC03-76SF00098 and in part by Conselho Nacional de Desenvolvimento
Cient\'{\i}fico e Tecnol\'ogico (CNPq), by Funda\c{c}\~ao de Amparo
\`a Pesquisa do Estado de S\~ao Paulo (FAPESP), and by Programa de
Apoio a N\'ucleos de Excel\^encia (PRONEX).
It was also supported by Spanish DGICYT under grants PB95-1077 and PB98-0693,
and by the European Union TMR network ERBFMRXCT960090.



\begin{table}
\begin{tabular}{||c||c|c|c|c|c|c||}
$M_H$ (GeV) & 100 & 110 & 120 & 130 & 140 & 150  \\
\hline
\hline
SM ($pp \to H \to \gamma\gamma$)  &
840 & 950 & 1040 & 950 & 715 & 560 \\
\hline
Background ($\gamma\gamma$)  &
40700 & 40700 & 29900 & 26300 & 22600 & 15300 \\
\hline
Background (jet--jet)  & 
1700 & 1300 & 1200 & 1200 & 1050 & 900 \\
\hline
Background ($\gamma$--jet)  & 
5000 & 3600 & 3200 & 2700 & 2250 & 1800 \\
\hline
${\cal S}/\protect\sqrt{{\cal B }}$  &
3.9     &   5.1  &  5.9  &  5.8  &  4.4  &  4.2 
\end{tabular}
\vskip 16pt
\caption{ Expected number of events for the Higgs production via
gluon--gluon fusion after cuts and efficiencies as given by the ATLAS
collaboration for an integrated luminosity of ${\cal L}=100$
fb$^{-1}$. We also exhibit the SM irreducible and reducible
backgrounds. We denoted by ${\cal S}/\protect\sqrt{{\cal B }}$ the
statistical significance of the SM Higgs signal.}
\label{gg:evt}
\end{table}


\begin{table}
\begin{tabular}{||c||c||}
$M_H$(GeV) &$(f_{BB}+f_{WW}-f_{BW})/\Lambda^2$ (TeV$^{-2}$)  \\
\hline
\hline
100 &($-$0.29,~0.38) or (2.1,~2.8) \\
\hline
110 &($-$0.23,~0.28) or (2.2,~2.8)  \\
\hline
120 &($-$0.21,~0.25) or (2.3,~2.8)\\ 
\hline
130 &($-$0.22,~0.27) or (2.5,~3.0)\\
\hline
140 &($-$0.28,~0.34) or (2.6,~3.3)  \\
\hline
150 &($-$0.35,~0.46) or (2.8,~3.6)  
\end{tabular}
\vskip 16pt
\caption{$95\%$ CL allowed ranges in TeV$^{-2}$ assuming that an
intermediate mass Higgs has been observed in gluon--gluon fusion
production with the SM rates for a luminosity of 100 fb$^{-1}$.}
\label{gg:h1}
\end{table}


\begin{table}
\begin{tabular}{||c||c||}
$M_H$(GeV) &$(f_{BB}+f_{WW}-f_{BW})/\Lambda^2$ (TeV$^{-2}$) \\
\hline
\hline
100 &(0.36,~2.1)  \\
\hline
110 &(0.49,~2.1)  \\
\hline
120 &(0.58,~2.1)  \\
\hline
130 &(0.59,~2.2)  \\
\hline
140 &(0.56,~2.4)  \\
\hline
150 &(0.52,~2.8)  \\
\end{tabular}
\vskip 16pt
\caption{$95\%$ CL sensitivity ranges in TeV$^{-2}$ coming from the
non observation of SM Higgs produced via gluon--gluon fusion.}
\label{gg:h2}
\end{table}


\begin{table}
\begin{tabular}{||c||c|c|c|c|c|c||}
$M_H$ (GeV) & 100 & 110 & 120 & 130 & 140 & 150  \\
\hline
\hline
SM $(pp \to VV \to H (\to \gamma\gamma) jj)$  &
61   &	79  &	93 & 93   & 78	 & 52  \\
SM background  &
69   &	71  & 67  & 64  & 58	 & 53  \\
${\cal S}/\protect\sqrt{{\cal B }}$  &
7.3     &   9.4  &  11.4  &  11.6  &  10.2  &  7.1
\end{tabular}
\vskip 16pt
\caption{Expected number of events for SM Higgs production via vector
boson fusion after including cuts and efficiencies for an integrated
luminosity of ${\cal L}=100$ fb$^{-1}$. We also present the SM
irreducible background and statistical significance of the signal.}
\label{vbf:evt}
\end{table}


\begin{table}
\begin{tabular}{||c||c||c||c||}
$M_H$(GeV) &$(f_{BB}+f_{WW}-f_{BW})/\Lambda^2$ (TeV$^{-2}$)  
& $f_{\Phi_2}/\Lambda^2$ (TeV$^{-2}$) &
$f_{all}/\Lambda^2$ (TeV$^{-2}$)  \\  
\hline
\hline
100 &($-$0.21,~0.26)  or (2.2,~2.7)  & ($-$2.2,~2.8) or (18,~23)
& ($-$0.20,~0.23) or  (2.6,~3.0) \\ 
\hline
110 &($-$0.18,~0.22) or (2.4,~2.8) & ($-$1.9,~2.2) or (24,~28)
& ($-$0.17,~0.20) or   (2.7,~3.1) \\ 
\hline
120 &($-$0.17,~0.19) or (2.5,~2.8) & ($-$1.7,~2.0) or (23,~27)  
& ($-$0.17,~0.19) or   (3.0,~3.4) \\ 
\hline
130 &($-$0.17,~0.20) or (2.6,~3.0) & ($-$1.7,~2.0) or (24,~28) 
& ($-$0.17,~0.19) or (3.1,~3.5) \\
\hline
140 &($-$0.21,~0.24) or (2.8,~3.2) & ($-$2.1,~2.3) or (37,~41)
& ($-$0.20,~0.23) or (3.3,~3.7) \\
\hline
150 &($-$0.29,~0.36) or(2.9,~3.6) &  ($-$2.8,~3.3)  or (32,~39)  
& ($-$0.29,~0.34) or (3.4,~4.0)    
\end{tabular}
\vskip 16pt
\caption{$95\%$ CL allowed ranges in TeV$^{-2}$ assuming that an
intermediate mass Higgs has been observed in the process $pp \to
qq^\prime VV \to j + j + H (\to \gamma\gamma)$ with the SM production
rate for an integrated luminosity of 100 fb$^{-1}$.}
\label{vbf:h1}
\end{table}


\begin{table}
\begin{tabular}{||c||c||c||c||}
$M_H$(GeV) &$(f_{BB}+f_{WW}-f_{BW})/\Lambda^2$ (TeV$^{-2}$) 
& $f_{\Phi_2}/\Lambda^2$ (TeV$^{-2}$) & $f_{all}/\Lambda^2$ (TeV$^{-2}$)  \\  
\hline
\hline
100 &(0.60,~1.9) & (7.2,~14.) 
& (0.67,~2.2) \\
\hline
110 &(0.70,~1.9) & (7.4,~18.) 
& (0.74,~2.1) \\ 
\hline
120 &(0.78,~1.9) & (8.5,~16.) 
& (1.0,~2.2) \\ 
\hline
130 &(0.84,~2.0) & (8.7,~17.) 
& (1.0,~2.3) \\
\hline
140 &(0.85,~2.1) & (7.6,~32.) 
& (0.99,~2.5) \\
\hline
150 &(0.79,~2.5) & (7.0,~29.)  
& (0.89,~2.8)  
\end{tabular}
\vskip 16pt
\caption{95\% CL sensitivity  ranges in TeV$^{-2}$ from vector boson
fusion production when no Higgs signal is observed in the VBF
channel.}
\label{vbf:h2}
\end{table}


\end{document}